\documentclass[
aps,
prd,
superscriptaddress,
nofootinbib,
floatfix]
{revtex4}
\usepackage{graphicx,
longtable,
color,
bm}
\begin{document}
\title{Updated bounds on the $(1,\,2)$ neutrino oscillation parameters after first JUNO results}
%
\author{        	Francesco~Capozzi}
\affiliation{   	Dipartimento di Scienze Fisiche e Chimiche, Universit\`a degli Studi dell'Aquila, 67100 L'Aquila, Italy}
\affiliation{		Istituto Nazionale di Fisica Nucleare (INFN), Laboratori Nazionali del Gran Sasso, 67100 Assergi (AQ), Italy}
\author{        	Eligio~Lisi}
\affiliation{   	Istituto Nazionale di Fisica Nucleare, Sezione di Bari, 
               		Via Orabona 4, 70126 Bari, Italy}
\author{        	Francesco~Marcone}
\affiliation{   	Dipartimento Interateneo di Fisica ``Michelangelo Merlin,'' 
               		Via Amendola 173, 70126 Bari, Italy}%
\affiliation{   	Istituto Nazionale di Fisica Nucleare, Sezione di Bari, 
               		Via Orabona 4, 70126 Bari, Italy}
\author{        	Antonio~Marrone}
\affiliation{   	Dipartimento Interateneo di Fisica ``Michelangelo Merlin,'' 
               		Via Amendola 173, 70126 Bari, Italy}%
\affiliation{   	Istituto Nazionale di Fisica Nucleare, Sezione di Bari, 
               		Via Orabona 4, 70126 Bari, Italy}
\author{        	Antonio~Palazzo}
\affiliation{   	Dipartimento Interateneo di Fisica ``Michelangelo Merlin,'' 
               		Via Amendola 173, 70126 Bari, Italy}%
\affiliation{   	Istituto Nazionale di Fisica Nucleare, Sezione di Bari, 
               		Via Orabona 4, 70126 Bari, Italy}
\begin{abstract}
\medskip
Within the standard $3\nu$  framework, we discuss updated bounds 
on the leading oscillation parameters related to the $(\nu_1,\,\nu_2)$ states, namely, the squared mass difference $\delta m^2=m^2_2-m^2_1$
and the mixing parameter $\sin^2\theta_{12}$.
A previous global analysis of 2024 oscillation data estimated $\delta m^2$ and $\sin^2\theta_{12}$ with 
fractional $1\sigma$ errors of about $2.3\%$ and $4.5\%$, respectively.
First we update the analysis by applying the latest  SNO+ constraints, 
that slightly shift the $(\delta m^2,\,\sin^2\theta_{12})$ best fits. 
Then we apply the constraints placed by the first JUNO results, that significantly reduce the uncertainties of both parameters. 
Our updated global bounds (as of 2025) can be summarized as:  \mbox{\boldmath$\delta m^2/10^{-5}{\rm eV}^2 = 7.48\pm 0.10$} and 
\mbox{\boldmath$\sin^2\theta_{12}=0.3085\pm0.0073$} (with correlation \mbox{\boldmath$\rho=-0.20$)}, corresponding to $1\sigma$ uncertainties as small as 1.3\% and 2.4\%, respectively.  We also comment on minor physical and statistical effects that, in the future, 
may contribute to lift the current mass-ordering degeneracy of
$(\delta m^2,\,\theta_{12})$ estimates.
\end{abstract}
\maketitle

\medskip
\section{Introduction}
\label{Sec:Intro}

Neutrino masses, mixing, and oscillations are at the forefront of modern particle physics \cite{ParticleDataGroup:2024cfk,PDG1}. 
In the standard three-neutrino ($3\nu$) framework, the neutrino flavor states $\nu_\alpha = (\nu_e,\,\nu_\mu,\,\nu_\tau)$ are mixed with neutrino mass states $\nu_i = (\nu_1,\,\nu_2,\,\nu_3)$ via a unitary mixing matrix $U_{\alpha i}$, parametrized in terms of three mixing angles  $(\theta_{12},\,\theta_{13},\,\theta_{23})$ and a CP-violating phase $\delta$. Observable oscillation phenomena are governed by (subsets of) the neutrino mixing parameters 
$\theta_{ij}$ and by the
squared mass differences $\Delta m^2_{ij}=m^2_i-m^2_j$ \cite{PDG1}. In particular, oscillations of solar and long-baseline (LBL) reactor neutrinos essentially probe the $U_{ei}$ mixing angles $(\theta_{12},\theta_{13})$ and the smallest $\delta m^2\equiv  \Delta m^2_{21} = m^2_2-m^2_1$, with $\theta_{13}$ strongly constrained by short-baseline (SBL) 
reactor neutrino experiments. The latter ones also probe an independent (larger) square mass difference, that is conventionally chosen as either 
$\Delta m^2_{31}$ or $\Delta m^2_{32}$, or a linear combination of them, such as $\Delta m^2_{ee} = 
\cos^2\theta_{12}\Delta m^2_{31}+\sin^2\theta_{12}\Delta m^2_{32}$ or $\Delta m^2 = (\Delta m^2_{31}+\Delta m^2_{32})/2$. The 
discrete parameter sign($\Delta m^2)=\pm 1$ distinguishes normal ($+1$) from inverted $(-1)$ mass ordering (NO and IO, respectively).
Atmospheric and long-baseline accelerator experiments can probe 
$(\Delta m^2,\,\theta_{23},\, \theta_{13}, \,\delta$), while having very weak sensitivity to $(\delta m^2,\,\theta_{12})$.
 
So far, oscillation experiments  have provided measurements of the squared mass differences $(\delta m^2,\,|\Delta m^2|$) and of the mixing angles $(\theta_{12},\,\theta_{13},\,\theta_{23})$, but not yet of $\delta$ or sign$(\Delta m^2)$ \cite{PDG1}. The status of known and unknown $3\nu$ parameters has been discussed 
in a recent global analysis of the neutrino oscillation data available in 2024 \cite{Capozzi:2025wyn}; see also 
\cite{Esteban:2024eli,deSalas:2020pgw}. In particular, Ref.~\cite{Capozzi:2025wyn} 
noted that the global fit accuracy of $|\Delta m^2|$ has reached, at face value, the subpercent level (0.8\%). 
In this context, the medium-baseline (MBL) reactor neutrino experiment at the Jiangmen Underground Neutrino Observatory (JUNO) 
\cite{JUNO:2015zny} stands on its 
own, being expected to reach subpercent precision at $1\sigma$ on three fundamental parameters $(|\Delta m^2|,\,\delta m^2,\,\sin^2\theta_{12})$ \cite{JUNO:2022mxj} and, in the long term, to determine also sign$(\Delta m^2)$ with a significance $>3\sigma$ \cite{JUNO:2024jaw}.
 Indeed, after starting data taking on 26 August 2025 \cite{JUNO:2025fpc},  JUNO has already measured 
$(\delta m^2,\,\sin^2\theta_{12})$ with a $1\sigma$ accuracy of $(1.6\%,\,2.8\%)$ in just 59.1 days \cite{JUNO:2025gmd}, 
to be compared with, e.g., the corresponding global fit accuracy of $(2.3\%,\,4.5\%)$ reported in \cite{Capozzi:2025wyn}. 
It thus makes sense to update the previous analysis \cite{Capozzi:2025wyn} for what concerns the parameters $(\delta m^2,\,\theta_{12})$, that govern 
solar and medium- or long-baseline reactor $\nu_e$ oscillations via $(\nu_1,\,\nu_2)$, when $(\Delta m^2,\,\theta_{13})$ are either fixed or marginalized away. 

For completeness, we take into account also 
the latest SNO+ long-baseline reactor data results 
\cite{SNO:2025chx}, that appeared just before the JUNO data release \cite{JUNO:2025gmd} 
but are comparatively less constraining.
We remark that, after including the current SNO+ and JUNO results, our update of $(\delta m^2,\,\theta_{12})$  does not  perceptibly alter the 
estimates of the other $3\nu$ oscillation parameters $(\Delta m^2,\,\theta_{23},\, \theta_{13}, \,\delta)$ or the relative likelihood of
NO and IO in the global oscillation fit, nor the upper limits on the 
absolute neutrino mass observables from $\beta$ decay ($m_\beta$), $0\nu\beta\beta$ decay ($m_{\beta\beta}$) and cosmology ($\Sigma$), that were previously  reported in \cite{Capozzi:2025wyn}. This motivates our focus on $(\delta m^2,\,\theta_{12})$ hereafter.

Our paper is structured as follows. In Sec.~\ref{External} we discuss the  
inclusion of the SNO+ and JUNO results as external constraints in the global fit.
In Sec.~\ref{Results} we report and comment the updated estimates of the $(\delta m^2,\,\sin^2\theta_{12})$ parameters in various graphical 
and numerical forms.
We summarize and conclude our work in Sec.~\ref{Conclude}.

\section{SNO+ and JUNO results as external constraints}
\label{External}

As in \cite{Capozzi:2025wyn}, we adopt a statistical $\chi^2$ approach and define $N_{\sigma}=\sqrt{\Delta \chi^2}$. By projecting the region allowed 
at $\Delta\chi^2 \leq N^2_\sigma$ onto one parameter $p$, one gets the range of $p$ at $N_\sigma$ standard deviations. In the context of this work,
when one dataset (e.g., SNO+ or JUNO) is separated---in a global $\chi^2$ fit---from all the other data, 
the latter are indicated as ``rest-of-the-world'' (RoW) data.   
The separate contributions of current SNO+ and JUNO results are implemented in a simple but effective way, as discussed below.  

\subsection{SNO+ constraints}
The SNO+ collaboration have reported their latest results \cite{SNO:2025chx}
in terms of a $\chi^2_{\rm SNO+}(\delta m^2,\,\sin^2\theta_{12})$ map for the two leading oscillation parameters. The subleading 
parameters $(\Delta m^2,\,\sin^2\theta_{13})$, taken from \cite{PDG1}, are marginalized away. 

In principle, since adding and projecting $\chi^2$ functions are noncommutative operations, a global fit to 
$(\delta m^2,\,\sin^2\theta_{12})$ should proceed as follows:
 $(i)$ Reconstruct the full-fledged analysis of SNO+ spectral data  
to obtain the four-parameter function $\chi^{2,\,\rm {full}}_{\rm SNO+}(\delta m^2,\,\Delta m^2,\,\sin^2\theta_{12},\,\sin^2\theta_{13})$; $(ii)$ add 
$\chi^{2,\,\rm {full}}_{\rm SNO+}$ to the rest-of-the-world function $\chi^2_{\rm RoW}$ (depending on all the $3\nu$ oscillation parameters);  and 
$(iii)$ marginalize away all parameters but $(\delta m^2,\,\sin^2\theta_{12})$.

In practice, since SNO+ data improve only weakly the RoW constraints on the $(\delta m^2,\,\sin^2\theta_{12})$
leading parameters (see \cite{SNO:2025chx} and the results in the next Section),  it is reasonable to neglect their (even weaker) effects on the subleading
$(\Delta m^2,\,\sin^2\theta_{13})$ parameters. 
We thus surmise that the impact of current SNO+ data on the 2024 global fit 
can be captured by simply adding---as an external constraint---the 
function $\chi^2_{\rm SNO+}(\delta m^2,\,\sin^2\theta_{12})$ from \cite{SNO:2025chx}  to the RoW function 
$\chi^2_{\rm RoW}(\delta m^2,\,\Delta m^2,\,\sin^2\theta_{12},\,\sin^2\theta_{13},\,\sin^2\theta_{23},\,\delta)$ from \cite{Capozzi:2025wyn}, and then marginalizing  the $(\Delta m^2,\,\sin^2\theta_{13},\,\sin^2\theta_{23},\,\delta)$ parameters, so as to obtain updated constraints in the
two-dimensional subspace charted by $(\delta m^2,\,\sin^2\theta_{12})$. 

\subsection{JUNO constraints}

The above arguments can also be applied to the first JUNO results \cite{JUNO:2025gmd}, 
although in part for different reasons. Concerning $\theta_{13}$, it is known that
the JUNO accuracy cannot be competitive with SBL reactor and LBL accelerator results, even after many years of operation \cite{JUNO:2022mxj}. Therefore,
as for SNO+, also JUNO is not expected to have a noticeable impact on RoW constraints on $\theta_{13}$. 

Concerning $|\Delta m^2|$, a similar statement is not so obvious: Indeed, after a couple of months of data taking, the expected 
JUNO accuracy should have already reached---in principle---a precision of $O(1\%)$ on $|\Delta m^2|$
\cite{JUNO:2022mxj}, competitive with current RoW constraints. In addition,
the $\Delta m^2_{ee}$ best fits in JUNO, for sign$(\Delta m^2)=\pm1$, are also expected to be displaced at $O(1\%)$, 
see e.g.~\cite{Capozzi:2025wyn} and references therein.  In any case, 
it is stated in \cite{JUNO:2025gmd} that: 
``Leaving $\Delta m^2_{31}$ unconstrained had a negligible impact on the results'' 
(on $\delta m^2$ and $\theta_{12}$ for NO), and also that: ``The results obtained for the inverted mass ordering
scenario are fully compatible'' (with those obtained for NO). Therefore, for some reasons 
to be understood (maybe statistical fluctuations or subtle systematics) 
the official analysis of first JUNO data is not yet appreciably sensitive to the magnitude and sign of $\Delta m^2$.

However, one expects that the JUNO sensitivity to $|\Delta m^2|$  will progressively emerge in the next months  of operation, 
together with a  sensitivity to sign($\Delta m^2$) in the next years.  
One will then need to reconstruct the two functions $\chi^2_{\rm JUNO}(\delta m^2,\,\Delta m^2,\,\theta_{12},\,\theta_{13})$ for NO and IO, add them to 
the corresponding functions $\chi^2_{\rm RoW}(\delta m^2,\,\Delta m^2,\,\sin^2\theta_{12},\,\sin^2\theta_{13},\,\sin^2\theta_{23},\,\delta)$, 
and eventually marginalize some oscillation parameters. A full-fledged analysis of JUNO spectral data, updating  the prospective ones 
reported in 
\cite{Capozzi:2013psa,Capozzi:2015bpa,Capozzi:2020cxm}, 
is in progress \cite{Progress}.

In conclusion, at present it seems acceptable 
to neglect the weak sensitivity of JUNO to both $\pm|\Delta m^2|$ and $\sin^2\theta_{13}$, 
and simply take the published bounds on ($\delta m^2,\,\sin^2\theta_{12}$) as external constraints in the global fit, 
similarly to the SNO+ case discussed above. 
We have checked that, to a very good approximation, the joint ($\delta m^2,\,\sin^2\theta_{12}$) bounds reported by JUNO in \cite{JUNO:2025gmd}  
correspond to a bivariate gaussian (and to a related $\chi^2_{\rm JUNO}$ function) with $1\sigma$ ranges 
given by $\delta m^2/10^{-5}{\rm \ eV}^2=7.500\pm 0.118$,  $\sin^2\theta_{12}=0.3092\pm 0.0087$, and correlation $\rho=-0.23$.
Concerning subleading mass-ordering effects versus future and more accurate data, see the comments at the end of Sec.~\ref{Results}.

\section{Updated \mbox{\boldmath$(\nu_1,\,\nu_2)$} parameters}
\label{Results}

\begin{figure}[t!]
\begin{minipage}[c]{0.85\textwidth}
\includegraphics[width=0.95\textwidth]{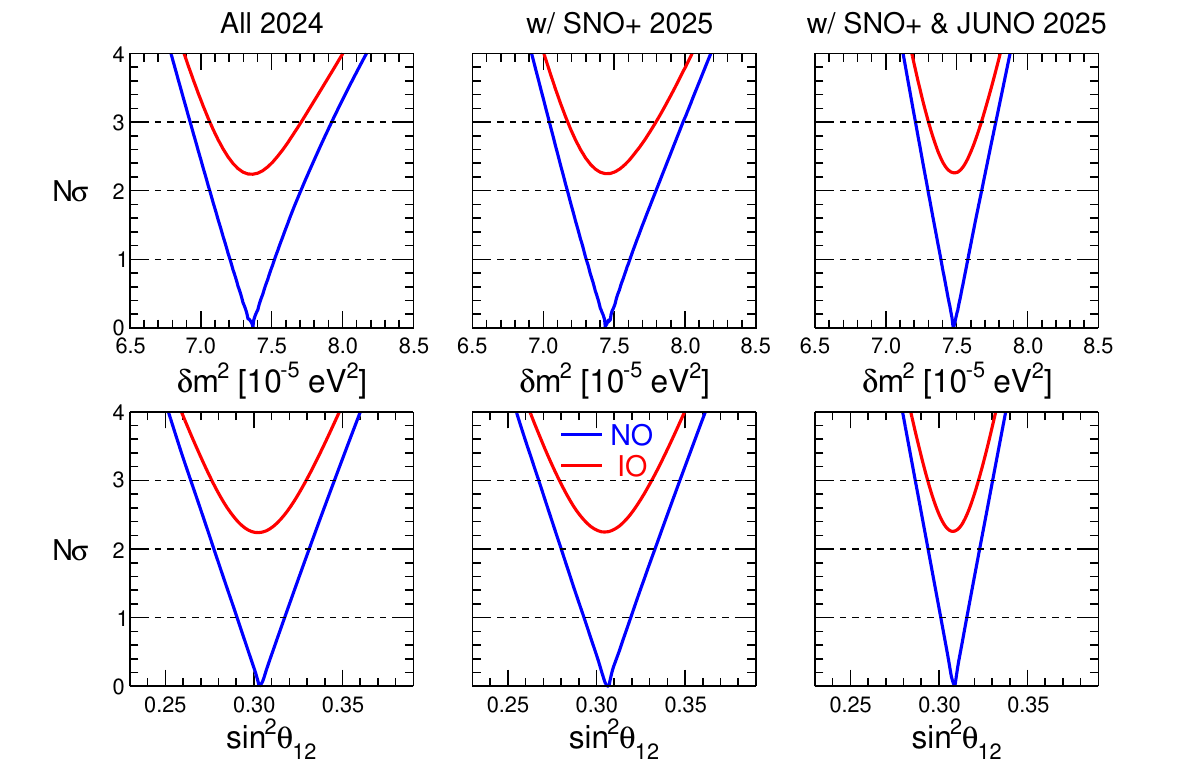}
\caption{\label{Fig_01}
\footnotesize Bounds at $N_\sigma$ standard deviations for $\delta m^2$ (upper panels) and $\sin^2\theta_{12}$ (lower panels)
for NO (blue curves) e IO (red curves) as obtained from 
a global data analysis of 2024 oscillation data \cite{Capozzi:2025wyn} (left panels), including---as detailed in this work---the latest 2025 SNO+ data \cite{SNO:2025chx} (middle panels), 
plus the first 2025 JUNO data \cite{JUNO:2025gmd} (right panels). The overall offset between the IO and NO curves is unchanged from \cite{Capozzi:2025wyn}.
See the text for details.  
} \end{minipage}
\end{figure}

Figure~\ref{Fig_01} shows the evolution of the $N_\sigma$ bounds on the $(\nu_1,\,\nu_2)$ oscillation parameters 
$\delta m^2$ (upper panels) and $\sin^2\theta_{12}$ (lower panels), starting from the ones published in \cite{Capozzi:2025wyn} (left panels)
and then adding the latest SNO+ constraints (middle panels) and the first JUNO constraints (right panels). The blue and red curves correspond 
to NO and IO, respectively, the latter being characterized by an offset $\Delta \chi^2_{\rm IO-NO}=5.0$ \cite{Capozzi:2025wyn}, unaltered by
SNO+ and JUNO data. It appears that SNO+ shifts upwards the best fits of ($\delta m^2,\,\sin^2\theta_{12}$) and slightly    
reduces their errors. JUNO increases the best fits a bit more and significantly reduces their uncertainties, that become almost
perfectly linear and symmetric.

Table~\ref{Tab:Synopsis} reports numerically the information in Fig.~\ref{Fig_01}, but assuming equiprobable NO and IO 
($\Delta \chi^2_{\rm IO-NO}=0$). There are no appreciable differences in the $\delta m^2$ and $\sin^2\theta_{12}$ bounds for NO and IO,
up to minor effects commented at the end of this section. From Fig.~\ref{Fig_01} and Table~\ref{Tab:Synopsis} we derive
our main result, namely, the estimated $1\sigma$ ranges and their correlation for  $\delta m^2$ and $\sin^2\theta_{12}$,
including all the available oscillation constraints as of 2025:
\begin{equation}
\label{Eq:Main}
{\rm All\ data\ 2025:\ }\left\{
\begin{array}{l}
\delta m^2/10^{-5}{\rm eV}^2 = 7.48\pm 0.10, \\ 
\sin^2\theta_{12}/10^{-1} = 3.085\pm0.073, \\
\rho=-0.20.
\end{array}
\right.
\end{equation}
These results are already dominated by JUNO, and will be even more so after future JUNO data releases. To a very good approximation,
the above errors scale linearly with $N_\sigma$. (We have averaged out tiny asymmetries and nonlinearities of positive and negative uncertainties.)
The value of the negative correlation is commented below.

\begin{table}[b]
\centering
\resizebox{.96\textwidth}{!}{\begin{minipage}{\textwidth}
\caption{\label{Tab:Synopsis} 
Best-fit values and allowed ranges at $N_\sigma=1$, 2, 3 for $\delta m^2$ and $\sin^2\theta_{12}$, 
as obtained in \cite{Capozzi:2025wyn} (all data 2024) 
and by adding the latest SNO+ data \cite{SNO:2025chx} and first JUNO results \cite{JUNO:2025gmd} as of 2025  (this work). 
The last column shows the formal  ``$1\sigma$ parameter accuracy,''  defined as 1/6 of the $\pm 3\sigma$ range, 
divided by the best-fit value (in percent).}
\begin{ruledtabular}
\begin{tabular}{lcccccc}
Global analysis [Ref.] & Parameter &  Best fit & $1\sigma$ range & $2\sigma$ range & $3\sigma$ range & ``$1\sigma$'' (\%) \\
\hline
All data 2024 & $\delta m^2/10^{-5}~\mathrm{eV}^2 $    & 	7.37 & 7.21 -- 7.52 & 7.06 -- 7.71 & 6.93 -- 7.93 & 2.3 \\
\cite{Capozzi:2025wyn} & $\sin^2 \theta_{12}/10^{-1}$ &  	3.03 & 2.91 -- 3.17 & 2.77 -- 3.31 & 2.64 -- 3.45 & 4.5 \\
\hline
w/ SNO+ 2025 & $\delta m^2/10^{-5}~\mathrm{eV}^2 $    &  	7.44 & 7.30 -- 7.61 & 7.17 -- 7.80 & 7.04 -- 7.99 & 2.1 \\
$[$This work$]$ & $\sin^2 \theta_{12}/10^{-1}$ &         	3.06 & 2.93 -- 3.19 & 2.80 -- 3.33 & 2.67 -- 3.47 & 4.4 \\
\hline
w/ SNO+ \& JUNO 2025 & $\delta m^2/10^{-5}~\mathrm{eV}^2 $&	7.48 & 7.39 -- 7.58 & 7.30 -- 7.68 & 7.21 -- 7.78 & 1.3 \\
$[$This work$]$  & $\sin^2 \theta_{12}/10^{-1}$ & 			3.085&3.010 -- 3.156&2.939 -- 3.230&2.866 -- 3.303& 2.4 \\
\end{tabular}
\end{ruledtabular}
\end{minipage}}
\end{table}

\begin{figure}[t!]
\begin{minipage}[c]{0.65\textwidth}
\includegraphics[width=0.99\textwidth]{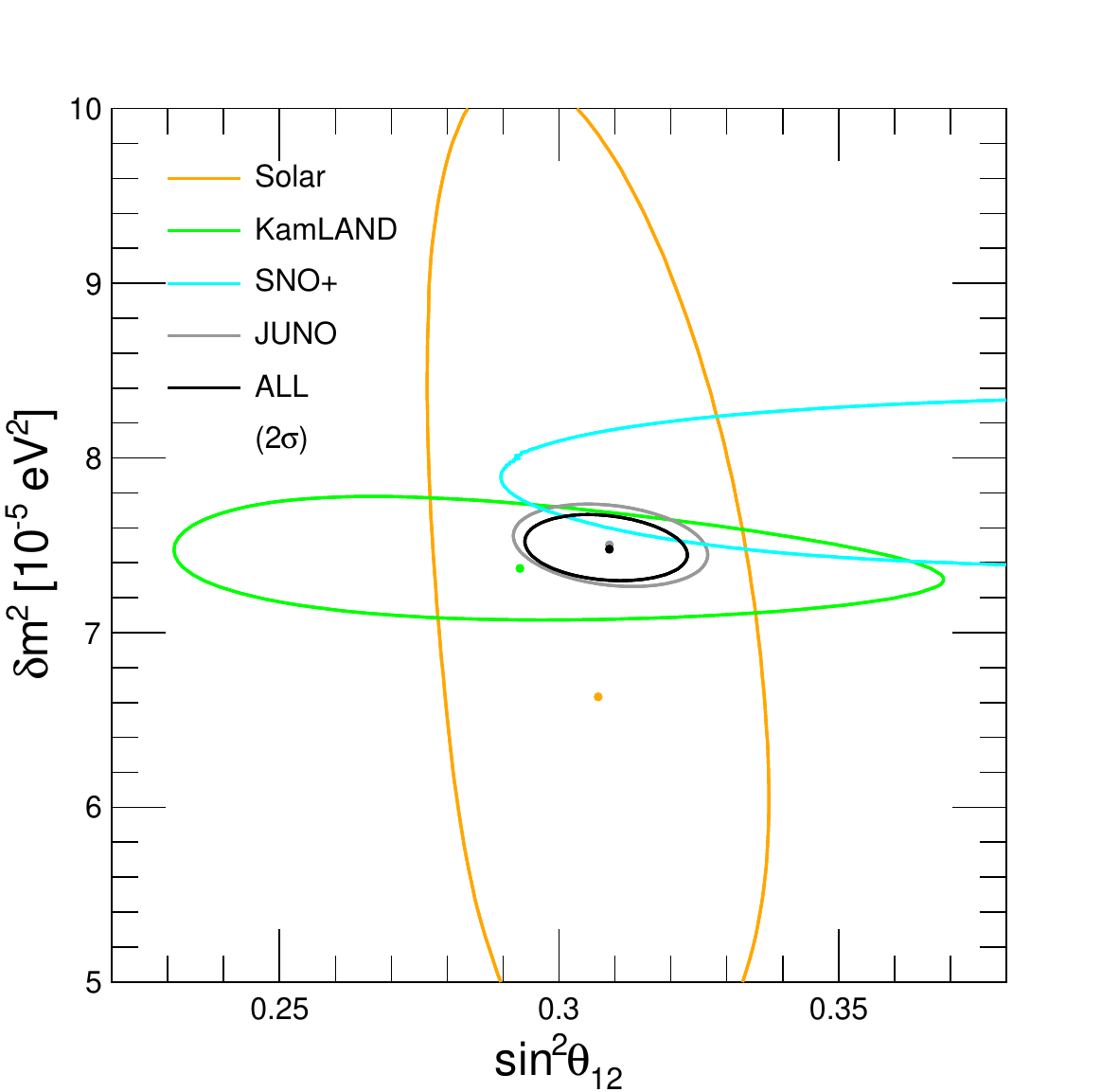}
\caption{\label{Fig_02}
\footnotesize Regions allowed at $N_\sigma=2$ ($\Delta\chi^2=4$) by separate and combined constraints
in the plane $(\delta m^2,\,\sin^2\theta_{12})$, distinguished by different colors. The same colors
are used to mark the corresponding best-fit points (dots). 
} \end{minipage}
\end{figure}

Figure~\ref{Fig_02} shows the results of our joint analysis of the
$(\delta m^2,\,\sin^2\theta_{12})$ parameters, all the other oscillation variables 
being marginalized away. Different colors in the legenda distinguish the
$N_\sigma=2$ contours and best fits (dots) for: solar and KamLAND LBL reactor
constraints (from the analysis in \cite{Capozzi:2025wyn}), SNO+ \cite{SNO:2025chx} and JUNO \cite{JUNO:2025gmd} constraints
(as detailed in Sec.~\ref{External}), and global 2025 constraints from all oscillation data (as discussed in this work).

Within the uncertainties, the separate 
experimental constraints in Fig.~\ref{Fig_02} appear to be in satisfactory agreement with each other, justifying
the global combination. The global fit for $(\delta m^2,\,\sin^2\theta_{12})$ 
is clearly dominated by the first JUNO results. It is then interesting to swap the combination sequence, 
starting from JUNO alone and adding rest-of-the-world data (where RoW includes SNO+). In this sequence (JUNO~$\to$~JUNO+RoW) one gets: $(i)$ a
downward shift
of the best-fit value by $-0.27\%$ for $\delta m^2$ and $-0.23\%$ for $\sin^2\theta_{12}$; $(ii)$ a reduction of the errors 
by a factor of about $\times 0.84$ for both parameters; 
and $(iii)$ a reduction of their correlation from $-0.23$ to $-0.20$.

\begin{figure}[t!]
\begin{minipage}[c]{0.62\textwidth}
\includegraphics[width=0.99\textwidth]{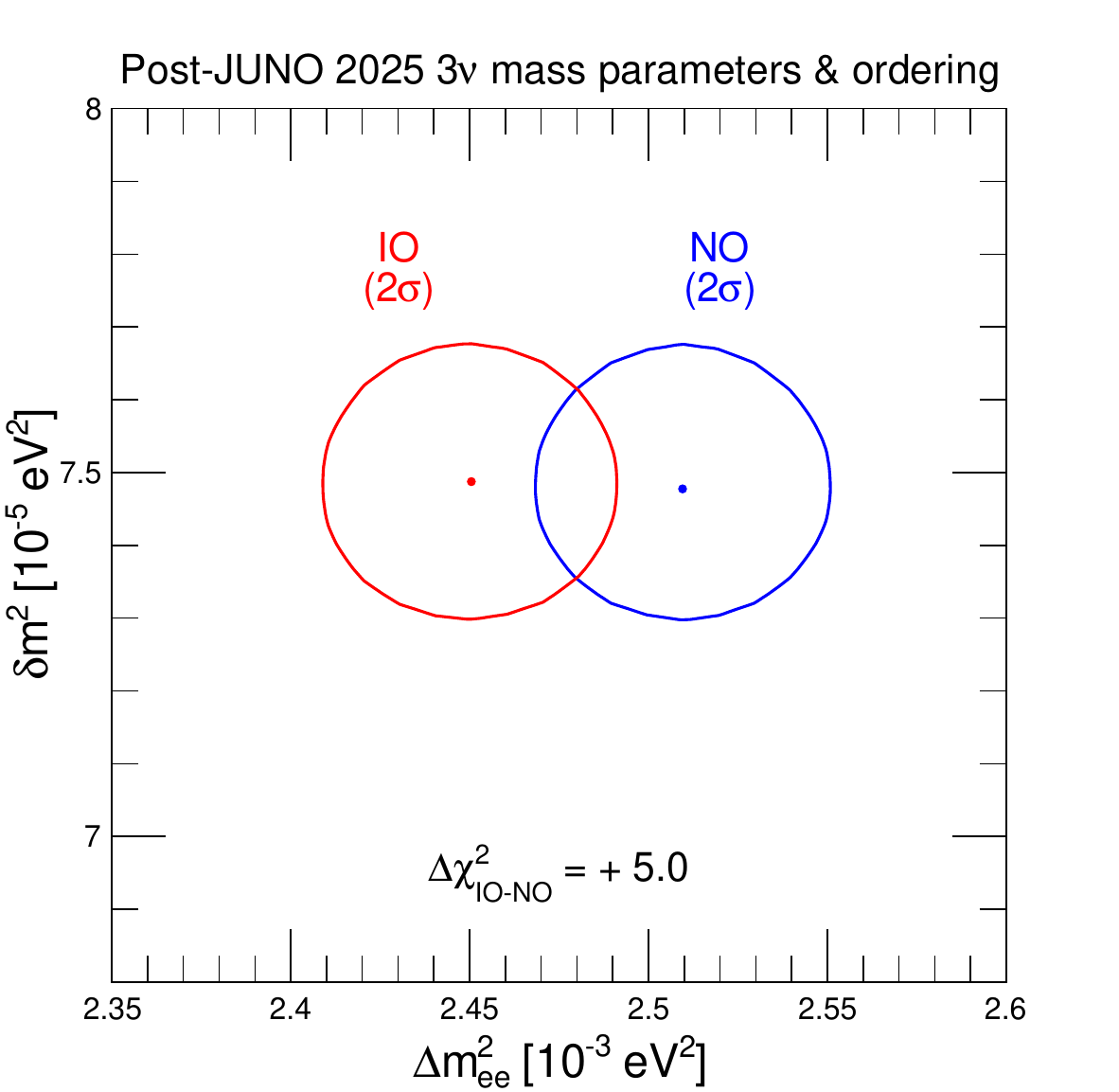}
\caption{\label{Fig_03}
\footnotesize Global $3\nu$ oscillation analysis (2025): Current $2\sigma$ bounds on the mass parameters 
$(\Delta m^2_{ee},\,\delta m^2)$ that govern the JUNO oscillation frequencies, 
for both NO (blue) and IO (red). The global $\Delta\chi^2$ offset between IO and NO is also reported.
} \end{minipage}
\end{figure}

Figure~\ref{Fig_03} shows the global $2\sigma$ contours (as of 2025) in the plane charted by the mass parameters 
$(\Delta m^2_{ee},\,\delta m^2)$ that govern the two oscillation frequencies observable in JUNO, for both NO and IO.
This figure updates the analogous pre-JUNO one shown in \cite{Capozzi:2025wyn} (see Fig.~7 therein). 
Further measurements of $\delta m^2$ and first determinations of $\Delta m^2_{ee}$ in JUNO (possibly in synergy
with other experiments) will help to progressively separate the NO and IO options
in the $(\Delta m^2_{ee},\,\delta m^2)$ plane, and eventually rule out one option,
if the underlying $3\nu$ framework is confirmed; 
see also the discussion and bibliography in \cite{Capozzi:2025wyn}, as well 
as representative cases of prospective JUNO data in \cite{FLASY25}.

Some final remarks are in order. Although we have shown the results of the $(\delta m^2,\,\sin^2\theta_{12})$ analysis independently of NO or IO,
minor differences between the two mass ordering options are either present or to be expected from the global fit to such parameters.
(E.g., in Fig.~\ref{Fig_03} one may appreciate a tiny vertical difference of the $\delta m^2$ best fit  in NO and IO.) 
The reasons for such minor differences are twofold.
 
One reason is physical: At high levels of accuracy, 
the leading approximation $\Delta m^2/\delta m^2 \to \infty$ for low-energy $\nu_e$ oscillations gets corrected by 
subleading $\pm |\Delta m^2|/\delta m^2$ effects in MBL reactor experiments
\cite{Petcov:2001sy}, as well as in other $\nu_e$ disappearance searches \cite{Fogli:2001wi}.  
At present, we have zeroed a priori the potential difference of SNO+ and of first JUNO reactor results in NO and IO, by using external constraints
independent of the mass ordering (see Sec.~\ref{External}). Although this approximation is largely justified for SNO+, 
our educated guess is that the current JUNO best fits might already be slightly sensitive to the mass ordering,
at least on the last (fourth) significant digit quoted for $\sin^2\theta_{12}$.   
We note that the $1\sigma$ range for IO, unreported in \cite{JUNO:2025gmd}, is declared to be ``fully compatible'' with the NO one 
(which does not necessarily mean identical).  On the other hand,
other subleading effects in $\nu_e$ oscillations are not zeroed a priori in our global analysis. In particular,
solar $\nu_e$ oscillations are slightly sensitive to $\pm|\Delta m^2|$ via 
small corrections to the effective mixing angles in matter at the neutrino production point \cite{Fogli:2001wi,Goswami:2004cn}.
 To a small extent, 
even SBL reactor neutrinos are slightly sensitive to the
mass ordering  \cite{Fogli:2001wi}.  Thus, 
for finite values of $\pm|\Delta m^2|/\delta m^2$, flipping the sign may induce tiny
phenomenological changes on $(\nu_1,\,\nu_2)$ oscillation parameters in $\nu_e$ disappearance experiments. Currently, 
we find that the subleading effects not zeroed in our analysis can 
change the best-fit estimates in Table~\ref{Tab:Synopsis} at the (few) permill level, 
namely, essentially in the fourth significant digit (or sometimes in the third, when the fourth is rounded off). 
Using four significant digits for $\sin^2\theta_{12}$ (both in JUNO \cite{JUNO:2025gmd} and in this work) 
allows to quote the $1\sigma$ error with two digits,  
but does not mean that the last digit is really determined. For definiteness,
our numerical results for $\sin^2\theta_{12}$ [as reported in Table~\ref{Tab:Synopsis} and Eq.~(\ref{Eq:Main})]
refer to NO, and generally coincide with the (unreported) IO results over the first three significant digits. 

A second reason for small mass-ordering differences in $(\delta m^2,\sin^2\theta_{12})$ estimates at (few) permill level 
is statistical. In a global fit, the update of one parameter value may affect (via correlations)
other parameters, even if the latter are not directly probed by the new input data.   
A well-known example 
is the estimate of $\Delta\chi^2_{\rm IO-NO}$ and of $(\sin^2\theta_{23},\,\delta)$, 
that are statistically affected---but not directly probed---by SBL reactor neutrino data, when combined with
atmospheric and LBL accelerator data; 
see, e.g., \cite{Capozzi:2025wyn} and references
therein. Similarly, the mass ordering correlates with $\Delta m^2$ and with $(\delta m^2,\,\sin^2\theta_{12})$ via JUNO:
flipping the sign of $\Delta m^2$ biases the JUNO best fits, at the percent level for $\Delta m^2$ and at the
few permill level for $(\delta m^2,\,\sin^2\theta_{12})$ \cite{Capozzi:2015bpa}. Conversely, updating the
global best-fit values of these oscillation parameters at such levels may change JUNO's estimate of $\Delta\chi^2_{\rm IO-NO}$ 
by a non-negligible amount \cite{Capozzi:2020cxm} and affect the IO$-$NO discrimination; 
see also \cite{JUNO:2022mxj,JUNO:2024jaw} and references therein.

In conclusion, small mass-ordering effects at the (few) permill level on $(\delta m^2,\,\sin^2\theta_{12})$---due 
to either subleading physics effects or the interplay among different constraints via statistical correlations---are not 
yet relevant, but may become so in future global fits including more accurate data, especially when JUNO will explore the
the subpercent precision range for one or more of its leading oscillation parameters.

\section{Summary and Conclusions}
\label{Conclude}

We have updated a previous global $3\nu$ analysis of 2024 neutrino data \cite{Capozzi:2025wyn} 
with regard to the leading $(\nu_1,\,\nu_2)$ oscillation parameters $(\delta m^2,\,\sin^2\theta_{12})$,  
by including as external constraints
the latest SNO+ \cite{SNO:2025chx} and first JUNO results \cite{JUNO:2025gmd} released in 2025. The updated global fit to such parameters
is dominated by JUNO, while the other data altogether produce a slight shift of the best-fit point and a reduction
of the JUNO errors by a factor of about $\times 0.84$. The main findings of our 2025 update are summarized
by Eq.~(\ref{Eq:Main}), in terms of global best-fit values and correlated $1\sigma$ errors for 
$(\delta m^2,\,\sin^2\theta_{12})$. To a very good approximation, such errors 
are symmetric and scale linearly with the number of standard deviations
$N_\sigma$. All the other $3\nu$ mass-mixing parameters  are not perceptibly changed, with respect to the allowed ranges or upper bounds 
estimated in \cite{Capozzi:2025wyn}.
We have presented graphical and numerical results that clarify the relative impact of SNO+ and JUNO data in the global fit to $(\delta m^2,\,\sin^2\theta_{12})$,
as well as their overall consistency with previous solar and LBL reactor data. We have also discussed minor variations
of $(\delta m^2,\,\sin^2\theta_{12})$ best-fit values and ranges between NO and IO, induced by either physical or statistical effects at the (few) permill level;
such effects are currently negligible, but  might be of some relevance in future analyses including tighter experimental constraints on the $3\nu$ mass-mixing parameters.

\acknowledgments

This work was partially supported by the research grant number 2022E2J4RK ``PANTHEON: Perspectives in Astroparticle and Neutrino THEory with Old and New messengers'' under the program PRIN 2022 funded by the Italian Ministero dell'Universit\`a e della Ricerca (MUR) and by the European Union -- Next Generation EU, as well as by the Theoretical Astroparticle
Physics (TAsP) initiative of the Istituto Nazionale di Fisica Nucleare (INFN). 
We thank W.~Giar\`e and A.~Melchiorri for previous collaboration on the analysis in \cite{Capozzi:2025wyn}.


{}

\end{document}